\PassOptionsToPackage{unicode}{hyperref}
\PassOptionsToPackage{hyphens}{url}
\documentclass[
  authoryear]{elsarticle}
\usepackage{amsmath,amssymb}
\usepackage{lmodern}
\usepackage{iftex}
\ifPDFTeX
  \usepackage[T1]{fontenc}
  \usepackage[utf8]{inputenc}
  \usepackage{textcomp} 
\else 
  \usepackage{unicode-math}
  \defaultfontfeatures{Scale=MatchLowercase}
  \defaultfontfeatures[\rmfamily]{Ligatures=TeX,Scale=1}
\fi
\IfFileExists{upquote.sty}{\usepackage{upquote}}{}
\IfFileExists{microtype.sty}{
  \usepackage[]{microtype}
  \UseMicrotypeSet[protrusion]{basicmath} 
}{}
\usepackage{xcolor}
\usepackage{graphicx}
\makeatletter
\def\maxwidth{\ifdim\Gin@nat@width>\linewidth\linewidth\else\Gin@nat@width\fi}
\def\maxheight{\ifdim\Gin@nat@height>\textheight\textheight\else\Gin@nat@height\fi}
\makeatother
\setkeys{Gin}{width=\maxwidth,height=\maxheight,keepaspectratio}
\makeatletter
\def\fps@figure{htbp}
\makeatother
\newlength{\cslhangindent}
\newlength{\csllabelwidth}
\newlength{\cslentryspacingunit} 
\newenvironment{CSLReferences}[2] 
 {
  \setlength{\parindent}{0pt}
  \ifodd #1
  \let\oldpar\par
  \def\par{\hangindent=\cslhangindent\oldpar}
  \fi
  \setlength{\parskip}{#2\cslentryspacingunit}
 }%
 {}
\usepackage{calc}

\usepackage{mathtools}
\usepackage{setspace}
\usepackage{amsmath}
\usepackage{scalerel}
\usepackage{stackengine}
\usepackage{booktabs}
\usepackage{longtable}
\usepackage{array}
\usepackage{multirow}
\usepackage{wrapfig}
\usepackage{float}
\usepackage{colortbl}
\usepackage{pdflscape}
\usepackage{tabu}
\usepackage{threeparttable}
\usepackage{threeparttablex}
\usepackage[normalem]{ulem}
\usepackage{makecell}
\usepackage{xcolor}
\ifLuaTeX
  \usepackage{selnolig}  
\fi
\IfFileExists{bookmark.sty}{\usepackage{bookmark}}{\usepackage{hyperref}}
\IfFileExists{xurl.sty}{\usepackage{xurl}}{} 
\hypersetup{
  pdfauthor={Bert van der Veen\^{}a},
  hidelinks,
  pdfcreator={LaTeX via pandoc}}

\author{Bert van der Veen\(^a\)}
\date{}

\begin{document}

\ead{bert_van_der_veen@hotmail.com}
\address[one]{Department of Mathematical Sciences, Norwegian University of Science and Technology, Trondheim, Norway}

\begin{abstract}
Principal component regression results in lack of fit when important dimensions are omitted, which cannot be assessed from the eigenvalues. I show that the PC-regression estimator can also suffer from increased variance relative to ordinary least squares in such cases. \noindent  \newline \textbf{keywords}:  dimension reduction, multicollinearity, PC-regression.
\end{abstract}
\title{A note on the variance in principal component regression}
\maketitle

\hypertarget{introduction}{%
\section{Introduction}\label{introduction}}

\noindent When there are many predictors in linear regression, or when
predictors are colinear, many researchers replace the predictors with
fewer principal components, derived from the predictor matrix, instead
(PCs, Pearson 1901). PCs have the benefit of being orthogonal, so that
including PCs has the tendency to stabilize computation (Hotelling 1957;
Jolliffe 1982).

Jolliffe (1982) noted that the original idea behind PC-regression was to
include all PCs, while authors usually omit PCs with small eigenvalues.
The arguments made against omitting PCs with small eigenvalues have
mostly been qualitative than quantitative (Jolliffe 1982; Artigue and
Smith 2019), but focuses on the premise that a small eigenvalue does not
reflect importance in explaining the response variable. Hadi and Ling
(1998) notes that the sum of squared errors obtained by PCR will never
be lower than that obtained by ordinary least squares (OLS). Næs and
Martens (1988) discusses the variance of the PC-regression estimator but
neglects to discuss how the PC-regression estimator for the residual
variance compares to the OLS estimator.

I develop various expressions for the variance of the PC-regression
estimator to show that the variance of the PC-regression estimator is
usually, but not always, smaller than that of the OLS estimator. This is
briefly demonstrated with the data from Christensen and Greene (1976).

\hypertarget{pc-regression}{%
\subsection{PC-regression}\label{pc-regression}}

\noindent When including \(d\) PCs in a regression, the matrix of
predictors \(\boldsymbol{X}\) with \(i = 1 \ldots n\) rows and
\(q = 1 \ldots p\) columns is first subjected to a Singular Value
Decomposition (SVD) to retrieve its left
\(\boldsymbol{U} = \boldsymbol{X}\boldsymbol{V} \boldsymbol{\Sigma}^{-1}\)
and right
\(\boldsymbol{V} = \boldsymbol{X}^\top\boldsymbol{U}\boldsymbol{\Sigma}^{-1}\)
singular vectors, where \(\boldsymbol{\Sigma}\) is a diagonal matrix of
singular values. In applied sciences, the left singular vectors take the
interpretation of gradients or latent variables, such as temperature,
climate, or genetic similarity in ecology and evolution, or types of
behavior such as aggressiveness, kindness, or passiveness in social
science. As such, the columns of \(\boldsymbol{U}\) represent a compound
of effects vaguely related to the original identities of the predictor
variables.

Let
\(\boldsymbol{X}_d = \boldsymbol{X} - \boldsymbol{U}_{k}\boldsymbol{\Sigma}_{k}\boldsymbol{V}_{k}^\top\)
denote the predictor variables ``observed'' with error and similarly for
\(\boldsymbol{X}_k\), or equivalently, the predictor matrix as
reconstructed from its SVD with the first \(d\) left singular vectors,
where \(\boldsymbol{\beta}_d\) and \(\boldsymbol{\beta}_k\) are the
(slope) parameters due to the first \(d\) and last \(k = p - d\) left
singular vectors of \(\boldsymbol{X}\), where usually \(d << p\). Then,
a linear regression can be formulated as: \begin{equation}
\begin{alignedat}{3}
\boldsymbol{y} &= \boldsymbol{X}\boldsymbol{\beta} &+ &\boldsymbol{\epsilon}\\
 &= \boldsymbol{X}_d\boldsymbol{\beta}_d + \boldsymbol{X}_k\boldsymbol{\beta}_k &+ &\boldsymbol{\epsilon}\\
&= \biggl(\boldsymbol{X} - \boldsymbol{U}_{k}\boldsymbol{\Sigma}_{k}\boldsymbol{V}_{k}^\top \biggr)\boldsymbol{\beta}_d + \biggl(\boldsymbol{X} - \boldsymbol{U}_{d}\boldsymbol{\Sigma}_{d}\boldsymbol{V}_{d}^\top \biggr)\boldsymbol{\beta}_k 
&+ &\boldsymbol{\epsilon}\\
&= \boldsymbol{X}\boldsymbol{\beta}_d &+ &\boldsymbol{\epsilon}_{d}.
\end{alignedat}
\label{one}
\end{equation} \noindent where \(\boldsymbol{\epsilon}\) is the error
from a regression with the predictor variables and
\(\boldsymbol{\epsilon}_d\) is the error from a regression with \(d\)
left singular vectors and similarly for \(\boldsymbol{\epsilon}_k\).
Next, let \(\boldsymbol{\beta}_{PC,d}\) represent the \(d\)-sized vector
of parameters for a PC-regression of the first \(d\) left singular
vectors, i.e., the parameters that correspond to the model:
\begin{equation}
\begin{alignedat}{4}
\boldsymbol{y} &= \boldsymbol{U}_{d}\boldsymbol{\beta}_{PC,d} &+ &\boldsymbol{\epsilon}_{d}\\
 &= \boldsymbol{X}_d\boldsymbol{\beta}_{d} &+ &\boldsymbol{\epsilon}_{d}.
\end{alignedat}
\label{two}
\end{equation} \noindent Consequently,
\(\boldsymbol{\beta} = \boldsymbol{V}\boldsymbol{\Sigma}^{-1}\boldsymbol{\beta}_{PC,d} + \boldsymbol{V}_k\boldsymbol{\Sigma}_k^{-1}\boldsymbol{\beta}_{PC,k} \equiv \boldsymbol{\beta}_d+\boldsymbol{\beta}_k\).
By noting that due to the orthogonality of the singular vectors
\(\boldsymbol{X}\boldsymbol{\beta}_d = \boldsymbol{X}_d\boldsymbol{\beta}_d\),
I arrive at the result
\(\boldsymbol{\epsilon}_{d} = \boldsymbol{\epsilon} + \boldsymbol{X}_k\boldsymbol{\beta}_k\),
where the second term in the error reflects the effect of ignoring the
last \(k\) left singular vectors.

\hypertarget{estimators-for-boldsymbolbeta-and-sigma2}{%
\subsection{\texorpdfstring{Estimators for \(\boldsymbol{\beta}\) and
\(\sigma^2\)}{Estimators for \textbackslash boldsymbol\{\textbackslash beta\} and \textbackslash sigma\^{}2}}\label{estimators-for-boldsymbolbeta-and-sigma2}}

\noindent Since
\(\hat{\boldsymbol{\beta}}_{PC,d} = \boldsymbol{U}_d^\top\boldsymbol{y}\),
the least squares estimator for \(\boldsymbol{\beta}\) admits to the
additive decomposition
\(\hat{\boldsymbol{\beta}} = (\boldsymbol{X}^\top\boldsymbol{X})^{-1}\boldsymbol{X}^\top\boldsymbol{y} = \boldsymbol{V}\boldsymbol{\Sigma}^{-1}\boldsymbol{U}^\top\boldsymbol{y} = \boldsymbol{V}_d\boldsymbol{\Sigma}_d^{-1}\boldsymbol{U}_d^\top\boldsymbol{y} + \boldsymbol{V}_k\boldsymbol{\Sigma}_k^{-1}\boldsymbol{U}_k^\top\boldsymbol{y} \equiv \hat{\boldsymbol{\beta}}_d+\hat{\boldsymbol{\beta}}_k\).

Next, assume
\(\text{var}(\boldsymbol{\epsilon}) = \sigma^2\textbf{I}_n\),
\(\text{var}(\boldsymbol{\epsilon}_d) = \sigma^2_d\textbf{I}_n\), and
\(\text{var}(\boldsymbol{\epsilon}_k) = \sigma^2_k\textbf{I}_n\). Let
\(RSS\) denote the residual sum of squares of a regression with the
predictors and \(RSS_d\) the residual sum of squares of a PC-regression
with \(\boldsymbol{H}_d = \boldsymbol{H}-\boldsymbol{H}_k\) the
PC-regression hat matrix. Let
\(\text{E}(\boldsymbol{y}) = \boldsymbol{X}(\boldsymbol{\beta}_d+\boldsymbol{\beta}_k)\)
so that the expectation of the residual sum of squared errors is:
\begin{align}
\begin{split}
\text{E}(RSS_d) &=\boldsymbol{y}^\top(\textbf{I}_n - \boldsymbol{H}_d)\boldsymbol{y}\\
&= \text{tr}[\{\textbf{I}_n-\boldsymbol{H}_d\}\{\text{E}(\boldsymbol{\epsilon}\boldsymbol{\epsilon}^\top) + \boldsymbol{X}(\boldsymbol{\beta}_d+\boldsymbol{\beta}_k)(\boldsymbol{\beta}_d+\boldsymbol{\beta}_k)^\top\boldsymbol{X}^\top\}]\\
&= \sigma^2(n-p) + \boldsymbol{\beta}_k^\top\boldsymbol{X}_k^\top\boldsymbol{X}_k\boldsymbol{\beta}_k,
\end{split}
\end{align} \noindent resulting in the estimator for the residual
variance in a PC-regression: \begin{equation}
\hat{\sigma}^2_{d} = \frac{\hat{\sigma}^2(n-p)}{n-d} + \frac{(\hat{\boldsymbol{\beta}}-\hat{\boldsymbol{\beta}}_d)^\top\boldsymbol{X}_k^\top\boldsymbol{X}_k(\hat{\boldsymbol{\beta}}-\hat{\boldsymbol{\beta}_d})}{n-d},
\label{varPC}
\end{equation} and the unbiased estimator for the residual variance of
linear regression \(\sigma^2\), based on the results from PC-regression,
is: \begin{equation}
\hat{\sigma}^2 = \frac{\hat{\sigma}^2_{d}(n-d)-\boldsymbol{y}^\top\boldsymbol{H}_{k}\boldsymbol{y}}{n-p}.
\label{sig}
\end{equation} \noindent The residual variance of a PC-regression can
alternatively be formulated as a function of the residual variance of a
regression for each separate dimension, due to the orthogonality of the
left singular vectors. Specifically,
\(\boldsymbol{y}^\top\boldsymbol{H}\boldsymbol{y} = \sum \limits^{p}_{q=1} \boldsymbol{y}^\top\boldsymbol{H}_{q} \boldsymbol{y}\),
where \(\boldsymbol{H}_{q}\) is the hat matrix of a regression that only
includes the \(q^{th}\) left singular vector with residual variance
\(\sigma^2_{q}\), so that: \begin{align}
\begin{split}
\hat{\sigma}^2_{d} &= \frac{\hat{\sigma}^2(n-p) + \boldsymbol{y}^\top\boldsymbol{y} -\hat{\sigma}_{k}^2(n-p+d)}{n-d}\\
 &= \frac{\hat{\sigma}^2(n-p) +  \boldsymbol{y}^\top\boldsymbol{y}k- (n-1)\sum \limits^{k}_{q=1}\hat{\sigma}^2_{q}}{n-d}\\
 &= \frac{(n-1)\sum \limits^{d}_{q=1}\hat{\sigma}^2_{q} -  \boldsymbol{y}^\top\boldsymbol{y}(d-1)}{n-d}.
\end{split}
\label{var3}
\end{align}

\hypertarget{bias-of-the-estimators}{%
\subsection{Bias of the estimators}\label{bias-of-the-estimators}}

\noindent The expression of the bias for \(\hat{\boldsymbol{\beta}}_d\)
as an estimator for \(\boldsymbol{\beta}\) is provided at the end of the
first section; for \(k\) omitted left singular vectors it is
\(\boldsymbol{\beta}_k\), so that \(\hat{\boldsymbol{\beta}}_k\) is an
estimate of the bias. The bias in the PC-regression estimator of the
residual variance is: \begin{equation}
\text{E}(\hat{\sigma}^2_{d}-\sigma^2) = (\frac{n-p}{n-d}-1)\sigma^2 + \frac{(\boldsymbol{\beta}-\boldsymbol{\beta}_d)^\top\boldsymbol{X}^\top\boldsymbol{X}(\boldsymbol{\beta}-\boldsymbol{\beta}_d)}{n-d},
\label{bias}
\end{equation} \noindent which shows that the bias is likely to increase
with the number of omitted left singular vectors \(k\), and that both
terms converge to zero as \(d \to p\), i.e., when the PC-regression
estimator coincides with the OLS estimator. However, note that the bias
induced depends on the importance of each omitted left singular vector
in explaining the response variable. As the first term of equation
\eqref{bias} is negative for \(n<d\), if none of the omitted left
singular vectors is relevant in explaining the response variable, the
residual variance is underestimated rather than overestimated by
PC-regression. However, overestimation can occur if the omitted left
singular vectors have a relationship with the response variable. As
\(n\to\infty\), the second term in equation \eqref{bias} dominates the
bias expression for the PC-regression estimator of the residual
variance. Note that if the omitted left singular vectors hold little
importance for the response variable, and when there are few included
left singular vectors, the bias in the residual variance estimate from
PC-regression will be negligible.

\hypertarget{variance-of-hatboldsymbolbeta-hatboldsymbolbeta_d-and-hatboldsymbolbeta_k}{%
\subsection{\texorpdfstring{Variance of \(\hat{\boldsymbol{\beta}}\),
\(\hat{\boldsymbol{\beta}}_d\), and
\(\hat{\boldsymbol{\beta}}_k\)}{Variance of \textbackslash hat\{\textbackslash boldsymbol\{\textbackslash beta\}\}, \textbackslash hat\{\textbackslash boldsymbol\{\textbackslash beta\}\}\_d, and \textbackslash hat\{\textbackslash boldsymbol\{\textbackslash beta\}\}\_k}}\label{variance-of-hatboldsymbolbeta-hatboldsymbolbeta_d-and-hatboldsymbolbeta_k}}

For the PC-regression estimator of the slope parameters I have so far
maintained the form: \begin{equation}
\hat{\boldsymbol{\beta}}_d \sim \mathcal{N}(\boldsymbol{V}_{d}\boldsymbol{\Sigma}^{-1}_{d}\boldsymbol{\beta}_{PC,d},\boldsymbol{V}_{d}\boldsymbol{\Sigma}^{-1}_{d}\boldsymbol{\Sigma}_{PC,d}\boldsymbol{\Sigma}^{-1}_{d}\boldsymbol{V}_{d}^\top),
\end{equation} \noindent or equivalently: \begin{equation}
\hat{\boldsymbol{\beta}}_d \sim \mathcal{N}(\boldsymbol{\beta}_d,\boldsymbol{V}_{d}\boldsymbol{\Sigma}^{-1}_{d}\boldsymbol{\Sigma}_{PC,d}\boldsymbol{\Sigma}^{-1}_{d}\boldsymbol{V}_{d}^\top).
\end{equation} \noindent However, the estimated covariance matrix of
\(\hat{\boldsymbol{\beta}}_d\) can alternatively be formulated in terms
of the covariance matrix of a regression with the predictors, and of a
regression with the remaining \(k\) left singular vectors. This will
facilitate a better understanding of changes in the variance of the
PC-regression estimator for different numbers left singular vectors in a
PC-regression. Let again
\(\boldsymbol{X}_d = \boldsymbol{X} - \boldsymbol{U}_{k}\boldsymbol{\Sigma}_{k}\boldsymbol{V}_{k}^\top\)
denote the predictors observed with error as included in a
PC-regression. Then, \begin{align}
\begin{split}
\text{var}(\hat{\boldsymbol{\beta}}_d) &= (\boldsymbol{X}_d^\top\boldsymbol{X}_d)^{-1}\sigma_{d}^2\\
&= \text{var}(\hat{\boldsymbol{\beta}})\boldsymbol{V}_{d}\boldsymbol{V}_{d}^\top\frac{\sigma_{d}^2}{\sigma^2},
\end{split}
\label{variance}
\end{align} where the diagonal of
\(\boldsymbol{V}_d\boldsymbol{V}_d^\top\) monotonically increases in the
number of left singular vectors in the regression \(d\). The fraction
\(\sigma^2_{d}/\sigma^2 > 1\) represents inflation in residual variance
of the PC-regression due to omission of the last \(k\) left singular
vectors (see appendix equation \eqref{A1} for a proof). This further
demonstrates that
\(\text{var}(\hat{\boldsymbol{\beta}})\boldsymbol{V}_{d}\boldsymbol{V}_{d}^\top \propto \text{var}(\hat{\boldsymbol{\beta}}_d)\),
with the benefit that \(\boldsymbol{V}_{d}\boldsymbol{V}_{d}^\top\) can
be computed without fitting a regression, but only with the matrix of
predictors and its SVD. The diagonal entries of the outer product of the
first \(d\) right singular vectors of \(\boldsymbol{X}\) represent the
proportional decrease in variance of the PC-regression estimator due to
omitting \(k\) left singular vectors. Consequently, if
\(\text{diag}(\boldsymbol{V}\boldsymbol{V}^\top\sigma^2_d/\sigma^2) >1\)
the variance of the PC-regression estimator for \(\boldsymbol{\beta}\)
is higher than that of the OLS estimator.

Finally, the variance of the PC-regression estimator can also be written
as a difference between the variance of the estimators of a regression
with the predictors, and that of the estimator for a regression of the
remaining \(k\) left singular vectors: \begin{align}
\begin{split}
\text{var}(\hat{\boldsymbol{\beta}}_d) &= \text{var}(\hat{\boldsymbol{\beta}})(\textbf{I}_p-\boldsymbol{V}_{k}\boldsymbol{V}_{k}^\top)\frac{\sigma_{d}^2}{\sigma^2}\\
&= \{\text{var}(\hat{\boldsymbol{\beta}})-\frac{\sigma^2}{\sigma_{k}^2}\text{var}(\hat{\boldsymbol{\beta}}_k)\}\frac{\sigma_{d}^2}{\sigma^2}.
\end{split}
\end{align}

\noindent Consequently,
\(\text{var}(\hat{\boldsymbol{\beta}}) = \text{var}(\hat{\boldsymbol{\beta}}_d)\sigma^2/\sigma^2_{d} + \text{var}(\hat{\boldsymbol{\beta}}_k)\sigma^2/\sigma_{k}^2\)
(see appendix equation \eqref{A2} for a proof). Here,
\(\sigma^2/\sigma^2_{k} < 1\) increases in \(k\) and converges to one as
\(k \to p\) and is zero at \(d = p\). To conclude, the PC-regression
estimator has sampling distribution
\(\hat{\boldsymbol{\beta}}_d \sim \mathcal{N}\{\boldsymbol{\beta}_d,\text{var}(\hat{\boldsymbol{\beta}})\boldsymbol{V}_{d}\boldsymbol{V}_{d}^\top\sigma^2_{d}/\sigma^2\}\),
or alternatively,
\(\hat{\boldsymbol{\beta}}_d \sim \mathcal{N}\{\boldsymbol{\beta}-\boldsymbol{\beta}_k,\text{var}(\hat{\boldsymbol{\beta}})\sigma_{d}^2/\sigma^2-\text{var}(\hat{\boldsymbol{\beta}}_k)\sigma_{d}^2/\sigma_{k}^2\}\).

\hypertarget{numerical-example}{%
\section{Numerical example}\label{numerical-example}}

\noindent That the variance of the PC-regression estimator can be higher
than that of the OLS estimator is straightforwardly demonstrated with an
example. The \texttt{Ecdat} \texttt{R}-package (Croissant and Graves
2022) includes the data from Christensen and Greene (1976) on the cost
of electricity production in the US during 1970. The dataset includes
information on the cost for producing electricity of \(n=158\)
companies. The seven predictor variables represent different components
of the production cost, including: the price and cost share (CS) for
labor, capital, and fuel, as well as the total output. The results of
regressions with the predictors, three PCs, and the remaining PCs are
presented in Table 1.

\setlength{\tabcolsep}{1pt}

\begin{table}

\caption{\label{tab:example}Parameter estimates with standard errors in brackets, for (PC-)regressions fitted to the elecricity data with $p=8$ predictors (including intercept), and $d=3$ (first three) and $k=5$ (last five) left singular vectors. Table entries for PC-regression with estimated standard errors larger than those estimated with OLS are bolded.}
\centering
\begin{tabu} to \linewidth {>{\raggedright}X>{\raggedright}X>{\raggedright}X>{\raggedright}X}
\toprule
  & $\hat{\boldsymbol{\beta}}$ & $\hat{\boldsymbol{\beta}}_d$ & $\hat{\boldsymbol{\beta}}_k$\\
\midrule
Intercept & 0.52 (0.015) & 0 (0) & \textbf{0.52 (0.031)}\\
Total output & 0.83 (0.015) & \textbf{0.084 (0.022)} & \textbf{0.74 (0.03)}\\
Wage rate & 0.037 (0.016) & \textbf{-0.041 (0.047)} & \textbf{0.078 (0.028)}\\
Labor CS & 0.04 (0.042) & \textbf{-0.17 (0.049)} & \textbf{0.21 (0.085)}\\
Capital price index & 0.03 (0.016) & \textbf{-0.005 (0.054)} & \textbf{0.035 (0.024)}\\
\addlinespace
Capital CS & 0.029 (0.045) & -0.038 (0.035) & \textbf{0.067 (0.093)}\\
Fuel price & 0.11 (0.018) & \textbf{0.015 (0.035)} & \textbf{0.094 (0.036)}\\
Fuel CS & -0.015 (0.061) & 0.14 (0.033) & \textbf{-0.16 (0.13)}\\
\bottomrule
\end{tabu}
\end{table}

\hypertarget{concluding-remarks}{%
\section{Concluding remarks}\label{concluding-remarks}}

\noindent It is well known that omitting important dimensions in
PC-regression results in lack of fit (Artigue and Smith 2019). In this
article I provide new quantitative arguments in support of existing
arguments against assessing the importance of dimensions in
PC-regression by their eigenvalues. The variance of the PC-regression
estimator for the regression slopes depends on the importance of the
omitted PCs in explaining the response variable, which cannot be
assessed from the eigenvalues of the PCA. If a PC is omitted that is
important in explaining the response variable, PC-regression will result
in lack of fit, the residual will be correlated with the predictors, and
as a consequence the residual variance can be overestimated.
Overestimation of the residual variance can lead to an increase in the
variance of the regression slope estimator, relative to the variance
estimated by OLS. By extension, if the omitted PC is not important in
explaining the response variable, PC-regression will provide a lower
estimate for the residual variance relative to the residual variance as
estimated by OLS, and the variance of the PC-regression estimator of the
regression slopes will be smaller than the variance of the OLS
estimator.

\hypertarget{acknowledgements}{%
\section{Acknowledgements}\label{acknowledgements}}

\noindent The writing of this note was motivated by the lack of nuance
on the variance of the PC-regression estimator in the associated
Wikipedia article (wikipedia.org/wiki/Principal\_component\_regression).
I would like to thank one anonymous reviewer, Erik Blystad Solbu and
Robert Brian O'Hara for comments on earlier drafts of the manuscript.

\appendix

\hypertarget{appendix}{%
\section{Appendix}\label{appendix}}

\numberwithin{equation}{section}
\setcounter{equation}{0}

\begin{align}
\begin{split}
\text{var}(\hat{\boldsymbol{\beta}}_d) &= (\boldsymbol{X}-\boldsymbol{U}_{k}\boldsymbol{\Sigma}_{k}\boldsymbol{X}_{\boldsymbol{V},k}^\top)^\top(\boldsymbol{X}-\boldsymbol{U}_{k}\boldsymbol{\Sigma}_{k}\boldsymbol{X}_{\boldsymbol{V},k}^\top)^{-1}\sigma_{d}^2\\
&= (\boldsymbol{V}_{d}\boldsymbol{\Sigma}_{d}\boldsymbol{U}_{d}^\top\boldsymbol{U}_{d}\boldsymbol{\Sigma}_{d}\boldsymbol{V}_{d}^\top)^{-1}\sigma_{d}^2\\
&= (\boldsymbol{V}_{d}\boldsymbol{\Sigma}_{d}\boldsymbol{\Sigma}_{d}\boldsymbol{V}_{d}^\top)^{-1}\sigma_{d}^2\\
&= \boldsymbol{V}_{d}\boldsymbol{\Sigma}_{d}^{-1}\boldsymbol{\Sigma}_{d}^{-1}\boldsymbol{V}_{d}^\top\sigma_{d}^2\\
&= \boldsymbol{V}_{d}\boldsymbol{V}_{d}^\top(\boldsymbol{X}^\top\boldsymbol{X})^{-1}\boldsymbol{V}_{d}\boldsymbol{V}_{d}^\top\sigma_{d}^2\\
&= (\boldsymbol{X}^\top\boldsymbol{X})^{-1}\boldsymbol{V}_{d}\boldsymbol{V}_{d}^\top\boldsymbol{V}_{d}\boldsymbol{V}_{d}^\top\sigma_{d}^2\\
&= (\boldsymbol{X}^\top\boldsymbol{X})^{-1}\boldsymbol{V}_{d}\boldsymbol{V}_{d}^\top\sigma_{d}^2\\
&= \text{var}(\hat{\boldsymbol{\beta}})\boldsymbol{V}_{d}\boldsymbol{V}_{d}^\top\frac{\sigma_{d}^2}{\sigma^2}.
\end{split}
\label{A1}
\end{align}

\begin{align}
\begin{split}
\text{var}(\hat{\boldsymbol{\beta}}_d) &= \text{var}(\hat{\boldsymbol{\beta}})(\textbf{I}_p-\boldsymbol{V}_{k}\boldsymbol{V}_{k}^\top)\frac{\sigma_{d}^2}{\sigma^2}\\
&= \text{var}(\hat{\boldsymbol{\beta}})(\textbf{I}_p-\boldsymbol{X}^\top\boldsymbol{X}\boldsymbol{V}_{k}\boldsymbol{\Sigma}_{k}^{-2}\boldsymbol{V}_{k}^\top)\frac{\sigma_{d}^2}{\sigma^2}\\
&= \{\text{var}(\hat{\boldsymbol{\beta}})-\text{var}(\hat{\boldsymbol{\beta}})\boldsymbol{X}^\top\boldsymbol{X}\boldsymbol{V}_{k}\boldsymbol{\Sigma}_{k}^{-2}\boldsymbol{V}_{k}^\top\}\frac{\sigma_{d}^2}{\sigma^2}\\
&= \{\text{var}(\hat{\boldsymbol{\beta}})-(\boldsymbol{X}^\top\boldsymbol{X})^{-1}\sigma^2\boldsymbol{X}^\top\boldsymbol{X}\boldsymbol{V}_{k}\boldsymbol{\Sigma}_{k}^{-2}\boldsymbol{V}_{k}^\top\}\frac{\sigma_{d}^2}{\sigma^2}\\
&= \{\text{var}(\hat{\boldsymbol{\beta}})-\sigma^2\boldsymbol{V}_{k}\boldsymbol{\Sigma}_{k}^{-2}\boldsymbol{V}_{k}^\top\}\frac{\sigma_{d}^2}{\sigma^2}\\
&= \{\text{var}(\hat{\boldsymbol{\beta}})-\frac{\sigma^2}{\sigma_{k}^2}\text{var}(\hat{\boldsymbol{\beta}}_k)\}\frac{\sigma_{d}^2}{\sigma^2}.
\end{split}
\label{A2}
\end{align}

\hypertarget{references}{%
\section{References}\label{references}}

\hypertarget{refs}{}
\begin{CSLReferences}{1}{0}
\leavevmode\vadjust pre{\hypertarget{ref-artiguePrincipalProblemPrincipal2019}{}}%
Artigue, Heidi, and Gary Smith. 2019. {``The Principal Problem with
Principal Components Regression.''} \emph{Cogent Mathematics \&
Statistics} 6 (1): 1622190.
\url{https://doi.org/10.1080/25742558.2019.1622190}.

\leavevmode\vadjust pre{\hypertarget{ref-christensen1976economies}{}}%
Christensen, Laurits R, and William H Greene. 1976. {``Economies of
Scale in {US} Electric Power Generation.''} \emph{Journal of Political
Economy} 84: 655--76.

\leavevmode\vadjust pre{\hypertarget{ref-croissantEcdatDataSets2022}{}}%
Croissant, Yves, and Spencer Graves. 2022. \emph{Ecdat: {Data} Sets for
Econometrics}. Manual. \url{https://CRAN.R-project.org/package=Ecdat}.

\leavevmode\vadjust pre{\hypertarget{ref-hadiCautionaryNotesUse1998}{}}%
Hadi, Ali S., and Robert F. Ling. 1998. {``Some {Cautionary Notes} on
the {Use} of {Principal Components Regression}.''} \emph{The American
Statistician} 52 (1): 15--19. \url{https://doi.org/10.2307/2685559}.

\leavevmode\vadjust pre{\hypertarget{ref-hotellingRelationsNewerMultivariate1957}{}}%
Hotelling, Harold. 1957. {``The {Relations} of the {Newer Multivariate
Statistical Methods} to {Factor Analysis}.''} \emph{British Journal of
Statistical Psychology} 10 (2): 69--79.
\url{https://doi.org/10.1111/j.2044-8317.1957.tb00179.x}.

\leavevmode\vadjust pre{\hypertarget{ref-jolliffeNoteUsePrincipal1982}{}}%
Jolliffe, Ian T. 1982. {``A {Note} on the {Use} of {Principal
Components} in {Regression}.''} \emph{Journal of the Royal Statistical
Society. Series C (Applied Statistics)} 31 (3): 300--303.
\url{https://doi.org/10.2307/2348005}.

\leavevmode\vadjust pre{\hypertarget{ref-naes1988principal}{}}%
Næs, Tormod, and Harald Martens. 1988. {``Principal Component Regression
in {NIR} Analysis: Viewpoints, Background Details and Selection of
Components.''} \emph{Journal of Chemometrics} 2 (2): 155--67.
\url{https://doi.org/10.1002/cem.1180020207}.

\leavevmode\vadjust pre{\hypertarget{ref-pearsonLIIILinesPlanes1901}{}}%
Pearson, Karl. 1901. {``{LIII}. {On} Lines and Planes of Closest Fit to
Systems of Points in Space.''} \emph{The London, Edinburgh, and Dublin
Philosophical Magazine and Journal of Science} 2 (11): 559--72.
\url{https://doi.org/10.1080/14786440109462720}.

\end{CSLReferences}

\end{document}